\documentclass[10pt,letterpaper,twocolumn]{article} 

\usepackage{ol2}
\usepackage[draft,implicit=false]{hyperref}
\usepackage{amsmath}

\begin{document}

\twocolumn[ 

\title{Dynamics of soliton explosions in passively mode-locked fiber lasers}


\author{Antoine F. J. Runge,$^*$ Neil G. R. Broderick and Miro Erkintalo}

\address{
Dodd-Walls Centre and the Department of Physics, The University of Auckland, Private Bag 92019, Auckland 1142, New Zealand.\\
$^*$Corresponding author: arun928@aucklanduni.ac.nz
}

\begin{abstract}
A soliton explosion is an instability  whereby a dissipative soliton undergoes a sudden structural collapse, but remarkably returns back to its original shape after a short
transient. We recently reported the first experimental observation of this effect in a fiber laser (A. F. J. Runge et al., Optica~2, 36 (2015)).
 Here, we expand on our initial work, presenting a more detailed experimental and numerical study of the characteristics and dynamics of soliton explosions in passively mode-locked fiber lasers. Specifically, we explore different cavity configurations and gain levels, observing and characterizing explosion events using spectral and temporal real-time single-shot
techniques. Our results highlight that the explosion characteristics observed in experiments depend critically on the position in the cavity where the output coupler is
located. Furthermore, we find that the frequency at which explosions occur can be controlled by adjusting the pump power. We also identify a new kind of ``partial'' explosion, where
strong spectral interference fringes appear on the pulse spectra, but a full collapse is avoided. Finally, we  perform numerical simulations based on a realistic iterative
cavity map, and obtain results that are in good agreement with experimental measurements. Careful analysis of the simulation results provide strong credence to the
interpretation that soliton explosions can be linked to a multi-pulsing instability.
\end{abstract}


 ] 

\section{Introduction}

Their ability to generate high-energy, short-duration pulses has made mode-locked fiber lasers highly desirable for a large range of applications \cite{fermann_review_2013,
 xu_review_2013}. But in addition to practical usage, ultrashort pulse lasers have also attracted significant interest in the field of nonlinear dynamics. This is because the
 complex interplay of linear and nonlinear effects, as well as gain and loss, can give rise to a rich diversity of different dynamical phenomena that can be systematically
 explored by controlling individual laser parameters (pump power, polarization, dispersion). Indeed, over the last decade, numerous nonlinear structures have been
 discovered and studied in mode-locked lasers~\cite{grelu_dissipative_2012}, including soliton molecules \cite{akhmediev_molecule_1997, grelu_molecule_2002,
 ortac_observation_2010, tsatourian_polarisation_2013}, soliton rains \cite{chouli_rain_2009, chouli_soliton_2010, bao_rain_2013, niang_rains_2014}, dissipative rogue
 waves \cite{kovalsky_extreme_2011, soto-crespo_dissipative_2011, zaviyalov_rogue_2012, lecaplain_dissipative_2012, lecaplain_dissipative_2013, runge_raman_2014,
 dudley_instabilities_2014, liu_rogue_2015}, and many other exotic localized structures~\cite{kelleher_chirped_2014, wabnitz_optical_2014, churkin_stochasticity_2015,
 chang_spiny_2015, chang_extreme_2015, chang_extreme2_2015}.

One of the most intriguing dissipative phenomena in mode-locked lasers is that of \textit{soliton explosion}. This corresponds to a process whereby a solitary pulse
circulating in the laser cavity experiences a transient structural collapse, but remarkably returns back to its previous state after a few roundtrips. The phenomenon was first
identified in numerical simulations of the complex cubic-quintic Ginzburg-Landau equation \cite{soto_pulsating_2000, akhmediev_explosion_2001}, and numerous studies
have subsequently used the same master equation to examine the effects of different parameters on the explosion dynamics
\cite{akhmediev_asymmetric_2004, latas_control_2010, latas_explosion_2011, cartes_noise_2012, cartes_symmetry_2014}. The first experimental observation was
reported in 2002 by Cundiff \textit{et al.}
in a mode-locked Ti:Sapphire laser \cite{cundiff_explosion_2002}. In particular, the authors used a diffraction grating and a photodetector-array to examine how the spectra
of the laser output pulses evolved over time. They observed clear signatures of explosion events, i.e.  the output spectrum occasionally experienced an abrupt structural
collapse, accompanied by a strong spectral blueshift. Until very recently, these results constituted the only experimental observation of soliton explosions.

We have recently reported the first observation of soliton explosions in a passively mode-locked \emph{fiber} laser \cite{runge_explosion_2015}. Specifically, we found
that explosion events manifest themselves when the laser parameters are carefully adjusted to lie between stable mode-locking \cite{erkintalo_gco_2012} and unstable
``noise-like'' (NL) pulse emission \cite{aguergaray_raman_2013}. We observed these events both in the spectral and in the temporal domains, using real-time shot-to-shot
measurement techniques. In the frequency domain, the explosions were found to manifest themselves through the collapse of the pulse's spectral structure and the
concomitant emission of strong red-shifted Raman components. After a few roundtrips, the soliton structure was found to spontaneously recover and to return to its
pre-explosion state, where it remained until another explosion occurred at a later time. In the time domain, each explosion event was found to be associated with an abrupt
temporal shift of the output pulse train. Our experimental results were supported by numerical simulations based on a set of generalized nonlinear Schr\"odinger
equations (GNLSE), which showed good agreement with experimental observations \cite{runge_explosion_2015}.

In our previous study \cite{runge_explosion_2015}, soliton explosions were only investigated for one particular set of laser parameters (pump power, cavity length). As a
consequence, many open questions remain that call for more detailed investigations. For example, do explosions only exist over a very narrow range of parameter, or can
they be observed for different experimental configurations? If the answer is positive, then how do the precise parameters influence the dynamics and characteristics of the
explosion events? Moreover, although our previous study alluded to a connection between multi-pulsing instabilities and soliton explosions, more detailed examinations are
required to fully understand the relationship.

In this article, we expand on our initial investigation, reporting a more comprehensive study of soliton explosion dynamics in passively mode-locked fibre lasers.
Specifically, we explore laser configurations with different cavity lengths and pump powers, systematically searching for and characterizing explosion events by
recording the shot-to-shot spectra of the output pulse trains using a dispersive Fourier transformation (DFT) technique
\cite{solli_dft_2008, solli_dft2_2012, wetzel_dft_2012, goda_dft_2013, godin_dft_2013, runge_coherence_2013}. We find that while explosions occur for  a
range of experimental parameters, their characteristics may depend sensitively on the particular configuration under study. We also investigate the impact of the
 laser mode-locker on the explosion characteristics, by adding a second output coupler in the cavity that provides an extra point of observation; our results show that the
 nonlinear transmission of the mode-locker has a very strong influence on the explosion energy characteristics. Finally, we examine in detail results from numerical
 simulations based on a set of nonlinear Schr\"odinger equations, obtaining important further insights into the origins of the observed soliton explosion events.
 In particular, through spectro-temporal analysis of the simulated laser output, we find that the explosions originate from the complex interaction between two partially
 overlapping pulses generated around the same central wavelength. Experiments performed with a significantly shortened laser cavity provide strong support for this
 hypothesis, allowing us to observe ``partial'' explosions that display measurable spectral inteference patterns characteristic of two closely-spaced pulses.

\section{Experimental setup and numerical model}

\begin{figure}[h]
\centering
\includegraphics[width=\columnwidth,clip = true]{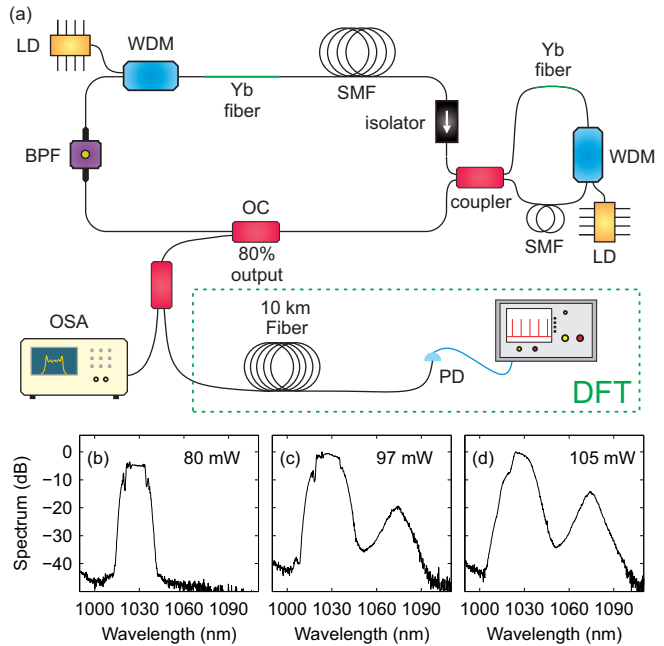}
\caption{\small{(a) Schematic of the cavity configuration and the DFT setup. LD, laser diode; WDM, wavelength division multiplexer; OC, output coupler; BPF, bandpass filter; SMF, single-mode fiber; PD, photodiode; OSA, optical spectrum analyzer. (b-d) Output spectrum of the laser operating in the mode-locking, explosion and noise-like regimes, respectively, with pump powers in the main loop as indicated. The NALM pump power is fixed to 150~mW.}}
\label{config}
\end{figure}

The core laser configuration examined in our study is similar to the one used in \cite{runge_explosion_2015, erkintalo_gco_2012} and it is schematically illustrated in
Fig.~\ref{config}(a). The laser is an all-normal dispersion~\cite{chong_all-normal_2006, chong_all-normal-dispersion_2007, renninger_giant_2008, kelleher_generation_2009, renninger_self-similar_2010, aguergaray_mode-locked_2012, aguergaray_120_2013}, all-polarization maintaining, all-fiber Yb-doped passively
mode-locked laser. The cavity follows a figure eight configuration, where the main loop incorporates a segment of Yb-doped fiber followed by a long segment of single-mode
fiber (SMF) and an optical isolator. The power is extracted out of the cavity using an 80~\% coupler, and at the end of each roundtrip, a spectral bandpass filter (1.7~nm
bandwidth) centered at 1030~nm is used to compensate for the large chirp accumulated over a single roundtrip. For mode-locking, we use a nonlinear amplifying loop
mirror (NALM)~\cite{fermann_nalm_1990}, which corresponds to the second loop in the cavity [right-hand side of Fig.~\ref{config}(a)]. The NALM is composed of an extra
patch of Yb-doped fiber and a segment of SMF. The doped fiber sections are pumped with two similar laser diodes emitting at 976~nm, and the laser self-starts when the
pump powers are correctly adjusted. Aside from the experiments reported in Section 6, the total length of the laser cavity $L_\mathrm{tot} \approx 140~\mathrm{m}$,
yielding a repetition rate of about 1.5~MHz. In that configuration, the total cavity dispersion is 3.5~ps$^2$ and the laser delivers pulse with a duration of 70~ps. Importantly for all the results that follow, the pump power in the NALM was fixed at 150~mW.

Below we present results obtained by fine tuning the pump power in the main loop when the laser is operating in the explosion regime, but we first recall that,
as shown in Fig~\ref{config}(b-d), coarse adjustment of the pump power allows significantly different laser operation regimes to be accessed
~\cite{runge_explosion_2015}.  For low pump power, the laser displays a stable output pulse train with a finely-structured spectrum [see Fig.~\ref{config}(b)]
\cite{erkintalo_gco_2012} while for high pump power the laser emits strongly fluctuating pulses that display noise-like (NL) pulse features [Fig.~\ref{config}(d)]
\cite{aguergaray_raman_2013, horowitz_noiselike_1997}. When operating in this regime, the output displays a second spectral component, generated by
stimulated Raman scattering (SRS), as described in \cite{aguergaray_raman_2013, north_raman-induced_2013}. (Note that, although in our configuration
strong SRS is indicative of unstable behaviour, in other laser designs it can lead to stable dual-wavelength operation~\cite{bednyakova_evolution_2013,
babin_multicolour_2014, kharenko_feedback-controlled_2015}.) Soliton explosions are observed when the pump power is adjusted to lie in the transition
zone between the stable and noise-like emission regimes~\cite{runge_explosion_2015}; first hints of this regime can be straightforwardly identified through
the averaged spectral characteristics that show features reminiscent of both the stable and the unstable regimes [see Fig.~\ref{config}(c)].

Because soliton explosions are a transient effect that occurs at megahertz repetition rates, they can be unequivocally captured and their dynamics examined only by
recording the roundtrip-to-roundtrip pulse evolution in real time. To achieve this, we implemented a dispersive Fourier transformation (DFT) technique that allows us
to record the roundtrip-to-roundtrip spectral evolution of the laser \cite{solli_dft_2008, solli_dft2_2012, wetzel_dft_2012, goda_dft_2013, godin_dft_2013,
runge_coherence_2013}. Concretely, when an optical pulse experience a sufficiently large dispersion, its temporal shape transforms to mimic its spectrum, and
each individual spectrum can then be recorded in the time domain using a fast photodiode and oscilloscope. This technique can be seen as the temporal equivalent
of the Fraunhofer diffraction. Our experimental DFT configuration is detailed in Fig.~\ref{config}(a). First, in order to avoid nonlinear shaping in the DFT setup, the average
power is decreased using an 1\% coupler. The attenuated output pulse train is then stretched in 10~km of fiber (group-velocity dispersion
$\beta_2=18.9\,{\rm ps^{2} \, km^{-1}}$) and the dispersed pulses are recorded using a 30~GHz photodiode and a 12~GHz real-time oscilloscope. With these parameters,
our DFT measurements have a spectral resolution of approximately 0.1~nm \cite{runge_coherence_2013}.

Finally, to gain more insight on the explosion formation, we have modelled the cavity using a fully realistic iterative cavity map detailed in \cite{runge_andi_2014}.
Briefly, the propagation of the optical pulse through each fiber segment (including those in the NALM) is simulated using a GNLSE, which includes stimulated and
spontaneous Raman scattering as well as higher dispersion \cite{dudley_rpm_2006}. The gains in the different doped-fiber sections are modeled for given pump
powers using an analytical three-level model \cite{Barnard-1994}. In our previous work we showed that this model is able to qualitatively describe the soliton explosion
dynamics, with numerical results showing good agreement with our experimental observations \cite{runge_explosion_2015}.

\section{Influence of the mode-locker}

\begin{figure}[t]
\centering
\includegraphics[width=\columnwidth,clip = true]{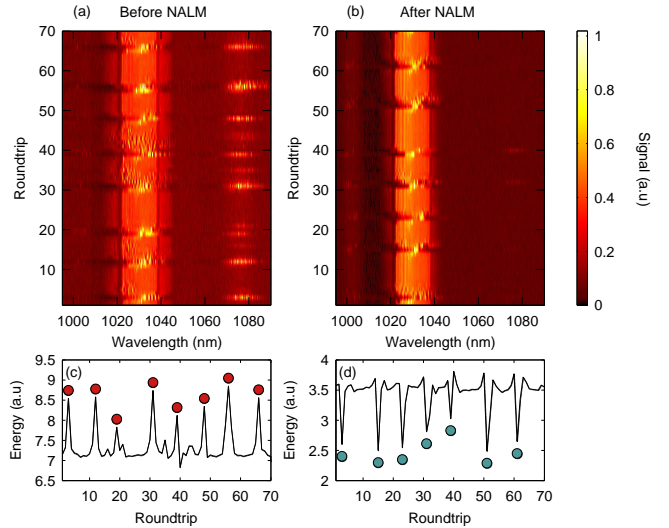}
\caption{\small{Experimental results, showing the shot-to-shot spectra of 70 consecutive pulses emitted by the laser when measured (a) before and (b) after the NALM mode-locker. (c) and (d) respectively show the corresponding roundtrip-by-roundtrip integrated pulse energy. For these measurements, the pump power was set at 100~mW in the main loop and 150~mW in the NALM. The solid circles mark the roundtrips corresponding to explosions.}}
\label{Fig2}
\end{figure}

\looseness=-1 We begin our study by examining the effect of the mode-locker on the observed explosion characteristics. In our previous study, we found that each explosion event was
associated with a significantly smaller energy than the quasi-stable pulses separating consecutive explosions \cite{runge_explosion_2015}. Surprisingly, this observation
contradicts prior numerical and experimental studies that have shown explosions to be associated with an abrupt energy \emph{increase} \cite{cundiff_explosion_2002,
soto_pulsating_2000}. However, in our laser design, the intracavity pulse exhibits considerable evolution over a single roundtrip~\cite{erkintalo_gco_2012}, suggesting
that the measured pulse characteristics can significantly depend on the point at which the laser output is extracted from. In our core configuration [see Fig.~\ref{config}(a)],
the laser output is extracted immediately after the NALM mode-locker, which has previously been shown to greatly influence the pulse energy statistics when the laser is
operating in the noise-like regime~\cite{runge_raman_2014}. Given that the explosion events resemble transient noise-like emission, it is reasonable to suspect that the
mode-locker may also influence the observed explosion features.

To  observe the impact of the mode-locker, we added a 1\% tap coupler between the isolator and the NALM. We then performed independent single-shot spectral
measurements both at the main output coupler and at the tap coupler, which allowed us to directly compare the explosion characteristics before and after the NALM.
Figure~\ref{Fig2} summarizes the results obtained from these measurements. In particular, the false colour density plots in Figs.~\ref{Fig2}(a) and (b) respectively
show the shot-to-shot spectra of 70 consecutive pulses measured before and after the NALM. We see that, both before and after the NALM, the main pulse centered
around 1030~nm displays explosion signatures similar to those reported in our previous work \cite{runge_explosion_2015}. Specifically, when an explosion occurs,
the pulse structure collapses with the output spectrum becoming much narrower and more intense. After a few roundtrips, the pulse structure returns back to its
previous state until another explosion occurs later on. We also notice, however, an important difference between the two measurements. In particular, before the
mode-locker the amplitudes of the Raman components generated during explosions are similar to those of the main pulses [Fig.~\ref{Fig2}(a)], while after the
NALM they are significantly attenuated [Fig.~\ref{Fig2}(b)].

Figures~\ref{Fig2}(c) and (d) show the roundtrip-by-roundtrip evolution of the total pulse energy before and after the NALM, respectively, obtained by integrating the
shot-to-shot spectra in Figs.~\ref{Fig2}(a) and (b) over the entire measured bandwidth. Before the NALM, the total energy abruptly \emph{increases} during each
explosion event, in agreement with behaviour reported in previous studies~\cite{cundiff_explosion_2002, soto_pulsating_2000}. In stark contrast, when characterised
after the NALM, at the main output coupler of our cavity, the energy \emph{reduces} during explosions as reported in our previous study~\cite{runge_explosion_2015}.
This discrepancy between the energy evolutions measured at the two different cavity positions can be readily understood by recalling that (i) explosions give rise to
Raman waves that are absent in between the events and (ii) the NALM in our laser tends to greatly attenuate the low power Raman components created by noise-like
pulses \cite{runge_raman_2014}. In particular, similarly to the dynamics of noise-like pulses, during each explosion a significant portion of the total pulse energy is converted
to Raman waves that are then reflected by the NALM, which leads to considerable reduction in total energy.

\section{Effect of the cavity gain on the explosion dynamics}

In this Section, we investigate the influence of the cavity gain on the dynamics of soliton explosions in our system. In particular, numerical studies have demonstrated
that the cavity gain can greatly influence the underlying dynamics and explosion characteristics \cite{akhmediev_explosion_2001, latas_control_2010}. This was
confirmed in experiments in a solid-state Ti:Sappire system, where gain was shown to allow for the frequency of explosion occurrence to be controlled
\cite{cundiff_explosion_2002}.

\begin{figure}[h]
\centering
\includegraphics[width=\columnwidth,clip = true]{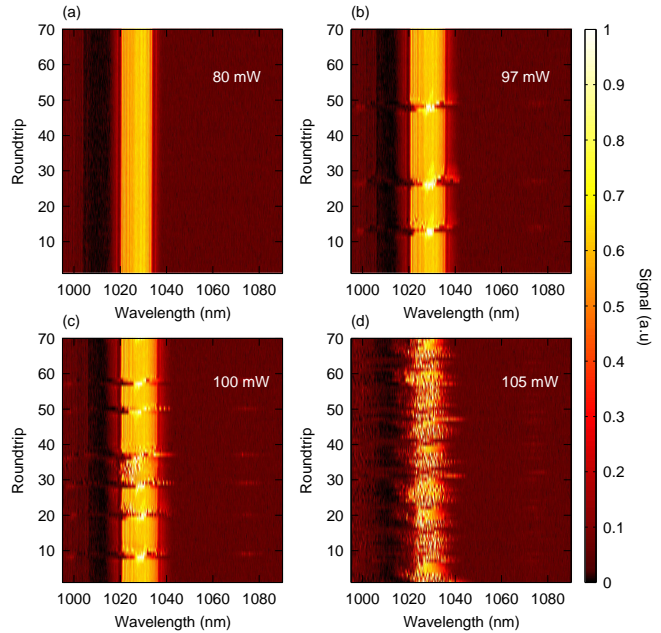}
\caption{\small{Experimentally measured single-shot spectra of 70 consecutive pulses with the laser operating in different regimes. In (a) the pump power in the main loop
was set to 80~mW and the laser emits a stable train of pulses. (b) For a pump power of 97~mW, we see three explosion events within the 70 roundtrip measurement. (c)
 Increasing the pump power to 100~mW leads to a higher number of explosions. (d) For 105~mW pump power the laser switches to the fully noise-like regime, exhibiting
 strong fluctuations from shot to shot.}}
\label{pump_vs_explo}
\end{figure}

In order to examine how cavity gain impacts on the explosion events in our fibre laser system, we first coarsely tune the laser pump power to lie close to the upper limit
of stable operation. We then increase the pump power in small steps, and for each value measure the roundtrip-by-roundtrip pulse spectra emitted by the laser. Typical
results from these experiments are shown in Fig.~\ref{pump_vs_explo}, where we plot 70 consecutive spectra output by the laser for pump power values ranging from
80~mW to 105~mW. When the pump power is set to 80~mW, the laser operates in the stable mode-locking regime and so the spectra of all output pulses are identical,
as shown in Fig.~\ref{pump_vs_explo}(a). When the pump power is slightly increased, the laser switches to the soliton explosion regime. In Fig.~\ref{pump_vs_explo}(b)
we show example spectra measured in this regime for 97~mW pump power, and we can clearly identify 3 explosions over the 70 recorded roundtrips. By slightly
increasing the pump power to 100~mW, we find that the number of explosions approximately doubles, as shown in Fig.~\ref{pump_vs_explo}(c). Finally, when the pump
is tuned to 105~mW, the laser switches to the NL regime and the output pulse train displays strong fluctuations from shot to shot \cite{runge_raman_2014}. These results
are particularly interesting, as they suggest -- even if just qualitatively -- that noise-like operation could potentially be understood as a limiting case of soliton explosions
occurring at each roundtrip. Analogously, the results shown in Fig. ~\ref{pump_vs_explo} hint that soliton explosions may be identified as transient noise-like pulses
that arise when the system's parameters momentarily cross the boundary between stable and unstable operation.

\begin{figure*}[t]
\centering
\includegraphics[width=\textwidth,clip = true]{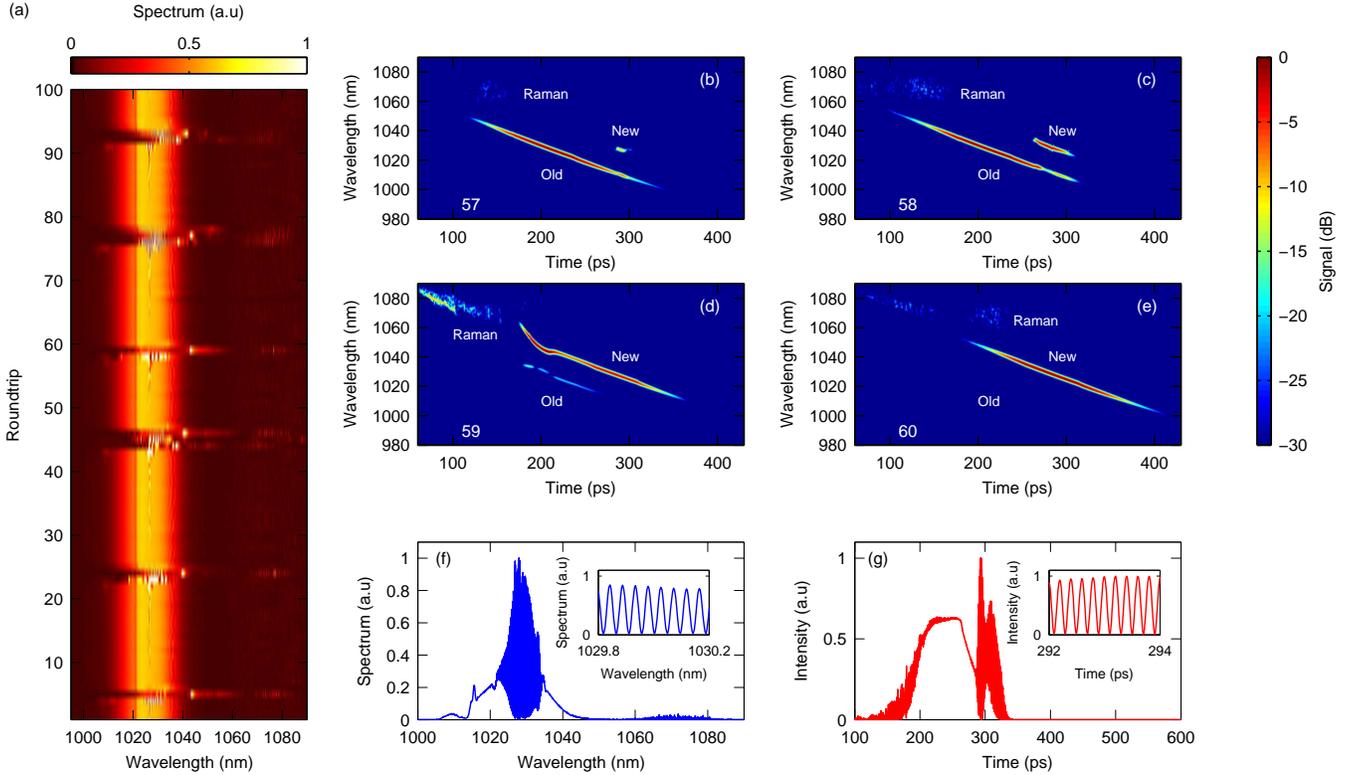}
\caption{\small{(a) Simulated output spectra over 100 consecutive roundtrips. (b-e) Simulated spectrograms of four consecutive roundtrips, with the corresponding
roundtrip numbers indicated on individual panels (bottom left). Simulated temporal (f) and spectral (g) output pulse profiles corresponding to the spectrogram shown
in (c). The insets in (f) and (g) show detailed zooms on the spectral and temporal interference patterns.}}
\label{simulation}
\end{figure*}

\section{Spectro-temporal explosion characteristics}

To gain more insights on the origins and dynamics of the explosion events, we now investigate in more detail their properties using numerical simulations. As discussed
above, the nonlinear cavity dynamics are modelled using the GNLSE that takes into account higher-order dispersion as well as spontaneous and stimulated Raman
scattering \cite{runge_andi_2014, dudley_rpm_2006}. Results from our numerical simulations show good qualitative agreement with experimental observations
\cite{runge_explosion_2015}, as seen in Fig.~\ref{simulation}(a), where we plot 100 consecutive simulated output spectra for parameters matching the experimental values.

We have found that the best way to visualise the results is through the use of a spectogram.  A spectrogram is a representation that allows for the temporal and
spectral contents of a signal to be simultaneously visualised (similar to a musical score), thereby permitting correlations between the two domains to be
straightforwardly identified~\cite{dudley_rpm_2006, genty_spectral_2002,
dudley_cross-correlation_2002, kelleher_generation_2009, erkintalo_experimental_2010, chapman_long_2010}. In Fig.~\ref{simulation}(b-e) we plot
spectrograms calculated for four consecutive numerically simulated output pulses, starting from the $57^\mathrm{th}$ roundtrip of the sequence shown in Fig.~\ref{simulation}(a).
For the roundtrip that just precedes the explosion onset [see Fig.~\ref{simulation}(b)], we can clearly identify two distinct pulses that are both centered at 1030~nm
but that are temporally separated by $\Delta \tau\sim 75~\mathrm{ps}$. Whilst the leading pulse (labelled ``old'') is broad and exhibits a large (linear) chirp, the
trailing one (labelled ``new'') has much smaller amplitude and a very narrow spectral width. However, since the small trailing pulse is centred at 1030~nm, it
survives the filter at the end of the roundtrip [see Fig.~\ref{config}(a)], allowing it to grow from one roundtrip to the next. Indeed, as seen in \ref{simulation}(c),
one roundtrip later the trailing pulse has grown much greater. Remarkably, in the subsequent roundtrip [see Fig.~\ref{simulation}(d)], which coincides with the
main explosion event, the trailing pulse has further grown to a magnitude that clearly now exceeds that of the critically diminished leading pulse. We also see
clearly how a very strong Raman component is emitted during the event. Finally, one roundtrip later [Fig.~\ref{simulation}(e)], the leading pulse has completely
vanished and only the second pulse persists, its characteristics similar to the original pulse aside from the evident temporal shift. These results very clearly
illustrate how the explosion events appear to be linked to the onset of a multi-pulsing instability.

In Figs.~\ref{simulation}(f) and (g) we plot in more detail the numerically simulated temporal and spectral profiles corresponding to the spectrogram shown in
Fig.~\ref{simulation}(c). We see that both profiles display very visible interference features; for the spectral profile fringes appear around the central wavelength
whilst in the temporal domain they reside at the trailing pulse edge. (The latter feature was already noted in~\cite{runge_explosion_2015}.) An analysis of the
spectrograms in Fig.~\ref{simulation}  enable us to intuitively explain these features. Indeed, the spectral fringes arise from the interference of the two temporally
separated pulses, with the fringe spacing related to the pulse temporal separation $\Delta \tau$ as $\Delta\lambda=\lambda^2/(\Delta\tau c)\sim 0.047~\mathrm{nm}$,
where $\lambda$ is the central wavelength and $c$ the speed of light. Also the temporal interference [see Fig.~\ref{simulation}(g)] is elucidated by inspection of the
corresponding spectrogram [Fig. \ref{simulation}(c)]. Specifically, the trailing ``new'' pulse overlaps temporally with the trailing edge of the ``old'' leading pulse.
But due to the strong frequency chirping of both pulses, their temporally overlapping parts are associated with a wavelength separation of
$\Delta\lambda \sim 17~\mathrm{nm}$, yielding a temporal interference pattern with fringe spacing $\Delta \tau = \lambda^2/(\Delta\lambda c)\sim 0.2~\mathrm{ps}$.

\section{Observation of partial explosions in a short laser cavity}

All the results presented above were obtained using a laser whose total roundtrip length $L_\mathrm{tot}\approx140~\mathrm{m}$, yielding a repetition rate of
about 1.5~MHz. Given that our previous investigations have shown that a similar laser architecture only supports unstable noise-like pulses for sufficiently long
cavities~\cite{aguergaray_raman_2013}, it is interesting to explore whether soliton explosions manifest themselves in lasers with higher repetition rates. To this
end, we have performed additional experiments after removing a significant portion of the SMF in between the main gain fiber and the optical isolator [see
Fig.~\ref{config}(a)], such that the laser repetition rate increases to about 4.5~MHz.

\begin{figure}[h]
\centering
\includegraphics[width=\columnwidth,clip = true]{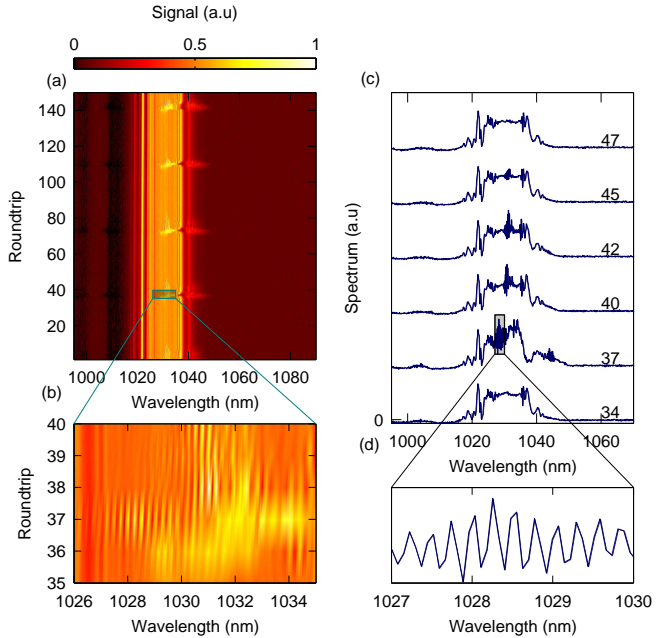}
\vskip 0mm
\caption{\small{Experimental results obtained for a shorter laser cavity with 4.5~MHz repetition rate. (a) Single-shot spectral measurement for 150 consecutive
roundtrips. (b) Zoom on the output spectra around the 37th roundtrip. (c) Individual examples of measured spectra around a particular explosion, with roundtrip
numbers as indicated. (d) Zoom over the spectral interference pattern of the 37th recorded roundtrip.}}
\label{5MHz_fringes}
\end{figure}

This shorter laser cavity is again observed to support three modes of operation. However, we find that, in the intermediate regime where explosion are to be
expected, the laser output here displays characteristics that are very different from the full explosions recorded for the longer laser. This is illustrated in
Fig.~\ref{5MHz_fringes}(a), where we show 150 consecutive experimentally measured spectra when the laser is operating in the regime between stable and
noise-like emission. In contrast to the longer laser cavity, the explosions here appear much weaker and they exhibit significant spectral asymmetry. In particular,
only the long-wavelength edges of the spectra are substantially influenced during the events captured in Fig.~\ref{5MHz_fringes}(a), whilst the short wavelength
edges remain almost totally unaffected. It is also noteworthy that the explosions here do not trigger the emission of Raman waves as they do in the longer cavity
\cite{runge_explosion_2015}. Due to their visibly weaker nature compared to previous observations~\cite{runge_explosion_2015}, we dub these
events ``partial explosions''.

\looseness=-1 Another intriguing feature of the short-cavity explosions is the fact that their spectra display clear interference fringes. This is highlighted in Fig.
\ref{5MHz_fringes}(b), where we show a zoom on the spectral evolution plot [Fig. \ref{5MHz_fringes}(a)] around a particular explosion event. For clarity, in
Fig.~\ref{5MHz_fringes}(c) we also show line plots of individual spectra at different roundtrips (indicated in the figure), and in Fig.~\ref{5MHz_fringes}(c) we
plot the interference pattern for a single realization in more detail. As explained in the previous Section, this spectral interference, that persists for several
roundtrips before and after the main explosion event, indicates that (at least) two temporally close pulses co-exist in the cavity. Accordingly, this observation
strongly supports the proposition that the explosions are intimately linked to a double-pulsing instability, as was discussed above [see also Fig.~\ref{simulation}(f)].
We note however that the fringe period $\Delta\lambda$, hence pulse temporal spacing $\Delta\tau$, observed in our measurements is very different from that
found in our simulations. Indeed, with $\Delta\lambda = 0.3$ nm, our measurements suggests that the two interfering pulses are  separated in time by only {12~ps}.

Before closing, we recall that the double-pulse dynamics during an explosion event causes the initial ``old'' pulse to switch into the final ``new'' pulse [see
Fig.~\ref{simulation}(b--e)], which gives rise to an observable temporal shift in the output pulse train~\cite{runge_explosion_2015}. Clearly, the magnitude of
this shift is indicative of the separation of the two competing pulses. For the longer cavity with 1.5~MHz repetition rate, we systematically find shifts in the range of
40~ps. Unfortunately, our current DFT setup has a resolution of about 0.1~nm, which is too large to resolve the interference between pulses spaced by more than
about 18~ps. We believe that it is this technical limitation alone that explains why spectral fringes are not observed in experiments involving the 1.5~MHz laser.

\section{Conclusion}
We have experimentally and numerically studied the dynamics of soliton explosions in a passively mode-locked fiber laser, and we have obtained several important
results. First, we  demonstrated that the measured explosion characteristics depend critically on the point in the laser cavity where the experimental observation is
carried out. In particular, we have shown that, if quantified before (after) the laser mode-locker, the explosion events can give rise to an overall increase (decrease)
in the total output energy. Second, we have explored how the explosion characteristics can be influenced by controlling the laser pump power. In agreement with
previous experimental studies of a Ti:Sapphire laser~\cite{cundiff_explosion_2002}, our measurements show that the number of explosions increase with cavity
gain. We have also studied in detail the spectral and temporal characteristics of explosion events using numerical simulations. By visualising the time-frequency
content of the laser output in terms of a spectrogram representation, we have clearly shown that the explosion dynamics are intimately linked to a double-pulsing
instability. Close inspection of the spectrograms has also allowed us to very intuitively explain particular interference features manifesting themselves in the
numerically simulated explosion envelopes. We have also experimented with a shorter laser cavity, observing a new regime of ``partial'' explosions whereby the
instability appears considerably weaker and only affects parts of the spectrum. Significantly, in this regime we have observed strong transient interference fringes,
which strongly corroborates the predicted double-pulsing origins of the explosion events. In addition to providing new insights into the dynamics of soliton
explosions in our particular laser configuration, we believe that our results will stimulate significant further research on similar instabilities in other configurations.

\section*{Funding Information}

\textbf{Funding.} Marsden fund of the royal society of New Zealand; the Finnish Cultural Foundation.


\end{document}